\documentclass[aps,prl,twocolumn,floatfix,superscriptaddress]{revtex4} 

\usepackage{graphicx}
\usepackage{amsmath}
\usepackage[amssymb]{SIunits}
\usepackage{color}

\usepackage[flushleft]{threeparttable}
\usepackage{epstopdf}
\usepackage{amssymb}
\usepackage{psfrag}

\renewcommand{\Re}{\operatorname{Re}}

\newcommand{\We}{\operatorname{We}}
\newcommand{\lam}{\operatorname{lam}}

\def\be{\begin{equation}}
\def\ee{\end{equation}}
\def\ba{\begin{eqnarray}}
\def\ea{\end{eqnarray}}

\begin{document}

\title{Bubble drag reduction requires large bubbles
}
\author{Ruben A. Verschoof} 
 
\affiliation{Department of Applied Physics and J. M. Burgers Center for Fluid Dynamics, University of Twente, P.O. Box 217, 7500 AE Enschede, The Netherlands}

\author{Roeland C.A. van der Veen} 
\affiliation{Department of Applied Physics and J. M. Burgers Center for Fluid Dynamics, University of Twente, P.O. Box 217, 7500 AE Enschede, The Netherlands}

\author{Chao Sun}
\email{chaosun@tsinghua.edu.cn}
\affiliation{Center for Combustion Energy and Department of Thermal Engineering, Tsinghua University, 100084 Beijing, China}

\affiliation{Department of Applied Physics and J. M. Burgers Center for Fluid Dynamics, University of Twente, P.O. Box 217, 7500 AE Enschede, The Netherlands}

\author{Detlef Lohse}
\email{d.lohse@utwente.nl}
\affiliation{Department of Applied Physics and J. M. Burgers Center for Fluid Dynamics, University of Twente, P.O. Box 217, 7500 AE Enschede, The Netherlands}
\affiliation{Max Planck Institute for Dynamics and Self-Organisation, 37077 G\"{o}ttingen, Germany}

\date{\today}

\begin{abstract} 
In the maritime industry, the injection of air bubbles into the turbulent boundary layer
under the ship hull is seen as  one of the most promising techniques
to reduce the overall fuel consumption. However, the exact mechanism behind bubble drag reduction is unknown. Here we show that bubble drag reduction in turbulent flow \textbf{\textit{dramatically}} depends on the bubble size.
By adding minute concentrations (6 ppm) of the surfactant Triton X-100 into otherwise completely unchanged
strongly turbulent Taylor-Couette
flow 
containing bubbles, we dramatically reduce the drag reduction from more than 40\% to 
about 4\%, corresponding to the trivial effect of the bubbles on the density and viscosity of the liquid. The reason for this striking behavior is that the addition of surfactants prevents bubble coalescence, leading to much smaller bubbles. Our result demonstrates that  bubble deformability is crucial for  bubble drag reduction in turbulent flow and opens
the door for an optimization of the process.

\end{abstract}

\maketitle

Theoretical, numerical and experimental studies on drag reduction (DR) of a solid
body moving in a turbulent flow have been performed for more than three decades \cite{cec10,mad84,mad85,lvov2005,murai2014,kumagai2015}. A few volume percent ($\leq 4\%$) of bubbles can reduce the overall drag up to 40\% and beyond
\cite{san06,deu04,sug04,ber05,ber07,elbing2008,gil13,elbing2013}. However, the exact physics behind this drag reduction mechanism is unknown,
thus hindering further progress and optimization, 
and even the dependence of the effect on the bubble size is controversial \cite{mer89,fer04,lu05}, though it 
is believed to be independent of the bubble size \cite{cec10}. 

In this Letter, we experimentally investigated the mechanism behind bubble drag reduction in a Taylor-Couette (TC) system, \textit{i.e.} the flow between two independently rotating coaxial cylinders. The TC system can be seen as 
``drosophila'' of physics of fluids, with many concepts in fluid dynamics being tested therewith,
ranging from instabilities, to pattern formation, to turbulence, see the reviews \cite{far14,gro16}.
Here we inject bubbles into the system, which due to the density difference to water
experience a centripetal force towards the inner cylinder, mimicking the upwards gravitational force 
acting on bubbles under a ship hull.

The experiments are performed in the Twente Turbulent Taylor-Couette facility (T$^3$C) \cite{gil11a}, with the inner one strongly rotating, corresponding
to very large Reynolds number of $\Re \sim 10^5 - 10^6$. The setup has an inner cylinder with a radius of $r_i = \unit{200}{\milli \meter}$ and an outer cylinder with a radius of $r_o = \unit{279}{\milli \meter}$, resulting in a radius ratio of $\eta = r_i/r_o = 0.716$. The inner cylinder rotates with a frequency up to $f_i=\unit{20}{\hertz}$, resulting in Reynolds numbers up to $\Re = 2\pi f_i r_i (r_o-r_i) /\nu_{\alpha} = 2 \times 10^6$, in which $\nu_{\alpha}$ is kinematic viscosity of water-bubble mixture. The outer cylinder is at rest. The cylinders have a height of $L = \unit{927}{\milli \meter}$, resulting in an aspect ratio of $\Gamma = L/(r_o-r_i) = 11.7$. The flow is cooled through both endplates to prevent viscous heating through the viscous dissipation.  
The torque $\tau$ is measured with  a co-axial torque transducer (Honeywell Hollow Reaction Torque Sensor 2404-1K, maximum capacity of \unit{115}{\newton \meter}),  mounted  
inside the middle section of the 
 inner cylinder, to avoid measurement uncertainties due to seals- and bearing friction and endplate effects. 
 Details are described in ref.\ \cite{gil11a}. 
 The gap between the cylinders is either fully 
filled with water ($T = \unit{20}{\degreecelsius}$) or, when measuring with bubbles,  partly filled with water 
($1-\alpha $).  The effective viscosity and density of a bubbly liquid can be approximated using $\rho_{\alpha}=\rho(1-\alpha)$ and 
the Einstein relation  \cite{gil13,ein1906}:  
$\nu_{\alpha} =\nu(1+ \frac{5}{2}\alpha)$, in which $\rho$ and $\nu$ are the density and the viscosity of the single phase liquid, and $\alpha$ is the \textit{global} volume fraction of air.  Air bubbles  form  over the entire cylinder height because of the
large turbulent fluctuations and the  high centripetal forces.

\begin{figure*}
\centering
\includegraphics[width=0.9\textwidth]{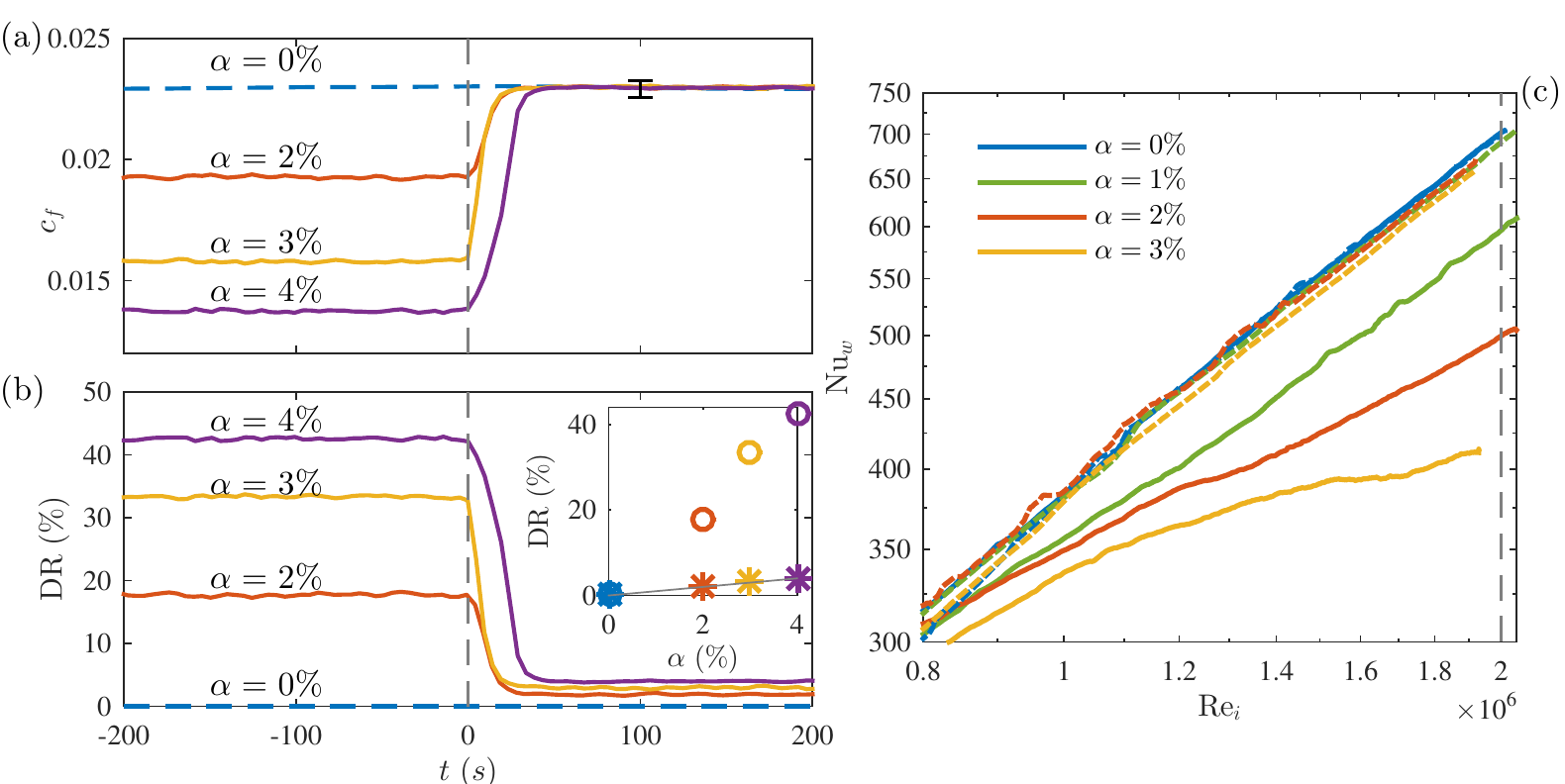}
\caption{
(a) Skin friction coefficient $c_f$ 
as function of time (for $f_i=\unit{20}{\hertz}$,
corresponding to 
 $\Re_i=2.0 \cdot 10^6$  at $\alpha = 0\%$) 
 for different gas volume fractions $\alpha$. At $t=0s$ the surfactant is injected, as indicated by the dashed vertical line. 
 We then observe a large jump in the measured friction coefficient 
 and within $\sim 20s$ 
 all curves
 overlap.
 (b)  
Drag reduction (DR) as  function of time. 
Nearly
 all DR is lost after injection of the surfactant at $t=0s$. 
 Inset: The averaged DR before (circles) and after (asterisks) 
 addition of the surfactant,
  as a function of the gas volume fraction $\alpha$. The thin line equals $\mathrm{DR}=\alpha$, 
  showing that after addition of the surfactant the small 
   residual DR is accounted for by the reduced density of the fluid mixture. 
(c) The dimensionless angular velocity transport  $Nu_\omega=\tau/\tau_{\lam}$, which is the angular velocity
transport ($\sim \tau $)  
divided by the angular velocity transport in the laminar and purely azimuthal case \cite{gro16}, 
as a function of the Reynolds  number $\Re$ 
for various  $\alpha=0\%,~1\%,~2\%,~3\%$, both with (dashed lines) and without (solid lines)
the surfactant Triton X-100. 
Figures (a) and (b) correspond to  $\Re = 2.0  \cdot 10^{6}$ (at $\alpha = 0\%$),
shown by the thin vertical line in the plot.  
}
\label{fig:jump}
\end{figure*}

The main result is seen in figure \ref{fig:jump}a,b, where we show the  drag coefficient $c_f(t)$ 
at $\Re_i = 2 \cdot 10^6$ as function
of time for four different bubble concentrations. It is calculated as 
$c_f  = \tau  / (L \rho_{\alpha} \nu_{\alpha}^2 \Re_i^{2} )$ 
 (see figure \ref{fig:jump}a)
from the measured 
required
torque $\tau (t)$ to keep the inner cylinder rotating at the fixed angular velocity $\omega_i$. 
While with  bubble volume concentration between 2\% and 4\% the drag is remarkably
reduced between  18\% - 43\%
as compared to the single phase flow case without bubble \cite{gil13} -- here the  percentage of  
drag reduction is expressed as 
$DR = (\tau^{with} - \tau ^{without})/\tau^{without}$
-- 
adding the surfactant Triton X-100 at $t=0s$ at a concentration of only 6 ppm reduces the drag reduction 
within $20s$ 
 (the time needed for Triton X to mix over the whole system) 
to the value corresponding to the volumetric gas concentration of 2\% - 4\%. 
The same holds for  weaker turbulence -- here we tested down to 
$\Re_i \approx  8 \cdot 10^5$ (see figure \ref{fig:jump}c) -- though for weaker turbulence the original drag reduction
effect through the bubbles is less pronounced.

\begin{figure*}
\centering
\includegraphics[width=0.8\textwidth]{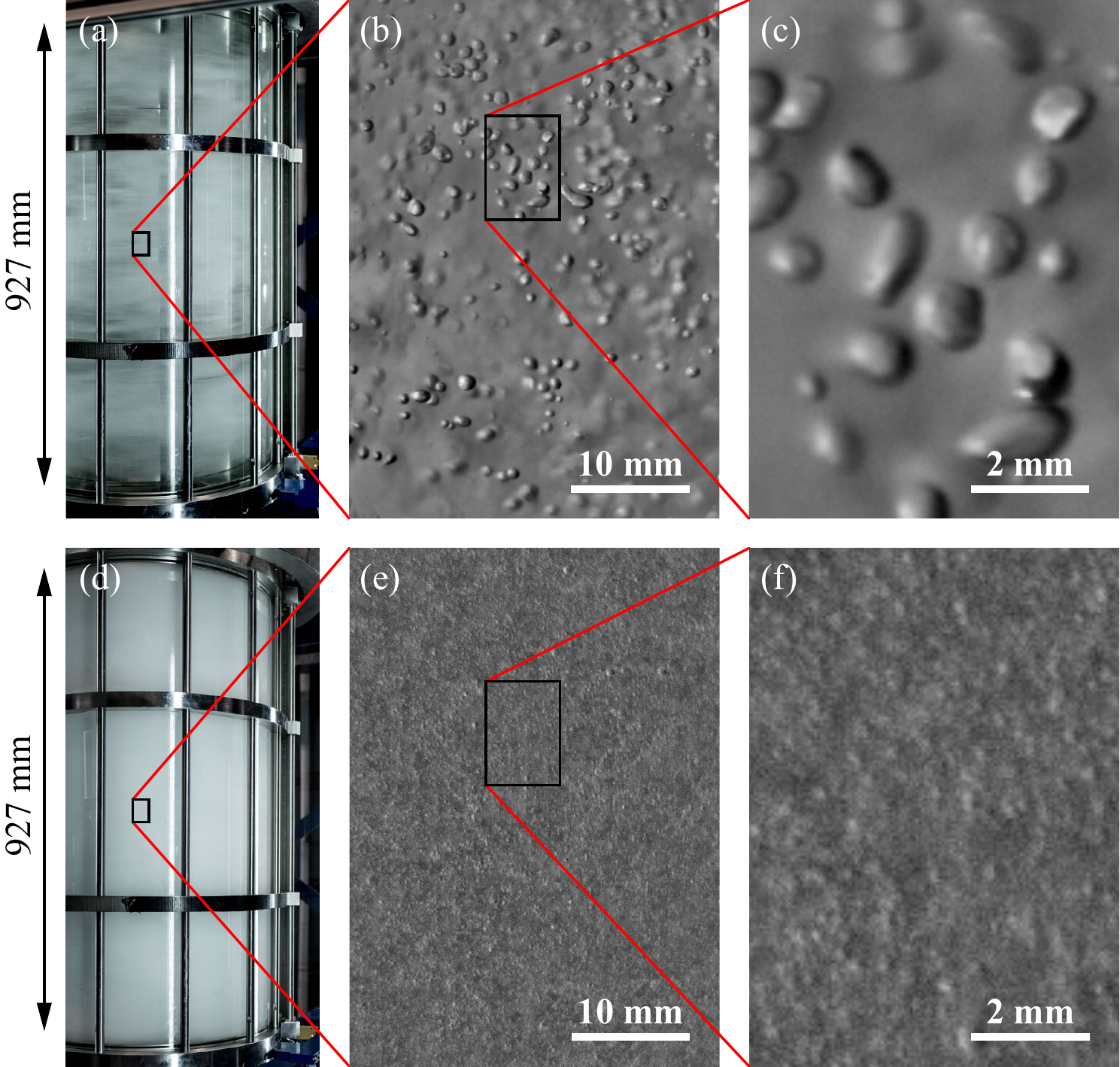}
\caption{
Snapshots of the bubbly turbulence ($\alpha =1\%$, $\Re_i = 2\cdot 10^6$) with increasing magnification (as shown by the scale bars).
In the first row no surfactants are present in the turbulent flow, whereas the second row shows
the (statistically stationary) situation after addition of 6 ppm Triton X-100. 
In the left photos the T$^3$C apparatus can be seen.
}
\label{fig:photos}
\end{figure*}

Figure \ref{fig:photos} shows snapshots of the bubbly turbulence at three different lengthscales (reflecting
the multiscale character of bubbly turbulence) without (upper row) and with (lower row) the addition of 
Triton X-100. It is seen that the addition of the surfactant dramatically changes the structure of the
turbulent dispersed bubbly flow, resulting in much smaller bubbles (with the same total volume concentration)
in the case with Triton X-100. 
The reason is that the surfactant suppresses bubble coalescence 
\cite{takagi2008,takagi2011}. 
Earlier studies noticed the role of the bubble Weber number in bubble drag reduction \cite{gil13,ber05,murai2014}. The Weber numbers $\We = \rho_{\alpha} u^{\prime 2 } D_{bubble}/\sigma $ before and after addition of Tritox X-100 are estimated as follows:
From fig.\ \ref{fig:photos}, we estimate that the equivalent 
bubble diameters are of order $D_{bubble,without}=O(1$ mm) for clean water, and $D_{bubble,with}=O(0.1$ mm) for water with Triton X-100, respectively. The surface tension between water and air is known for clean water, i.e. $\sigma_{without}=$ 73~mN/m at room temperatures. After the addition of 6ppm Triton X-100 (equivalent to $5\cdot 10^{-5}$ mol/L), the surface tension lowers to  $\sigma_{with}=$ 40~mN/m \cite{gob97}. The velocity fluctuations are impossible to measure after the addition of the surfactant, the flow is too dense to be optically accessible. We know that \textit{without bubbles}, $u^{\prime} _{\theta}\approx 0.03 \omega_i r_i$ \cite{gil13} in the bulk of the flow, and that this ratio is constant over a large range of Reynolds numbers, as long as the flow is fully turbulent \cite{hui12}. Furthermore, it has been shown that this ratio does not change much after adding a few percent of mm-sized bubbles \cite{gil13}. For a rotation rate of 20 Hz, we calculate that $u^{\prime}=$ 0.76 m/s. We assume that this is a reasonable measure for the fluctuations in our bubbly flow. For lower Reynolds numbers, the velocity fluctuations become smaller, resulting in lower Weber numbers.

From the figures we 
 estimate the corresponding Weber numbers 
in the two cases as $\We_{without} \approx 10 $  and  $\We_{with} \approx 1 $, implying that prior to injection of the
surfactant the bubbles can deform (as indeed seen from the figures \ref{fig:photos}b,c), whereas 
this is not possible after Triton X-100 was added (which is consistent with figures \ref{fig:photos}e,f). 
As shown in figure\ \ref{fig:jump}c, 
drag reduction is less pronounced at lower Reynolds numbers. The physical reason for this trend is that the
Weber number of the bubbles decreases when reducing the Reynolds number.

Our findings  give strong evidence that the bubble deformability is crucial in the drag reduction mechanism,
as  already speculated in refs.\ \cite{lu05,ber07,gil13}, but disputed by other authors. We  note that both the shape 
change of the bubble and the bubble coating by the surfactant
will also modify the lift force coefficient of the lift acting on the bubble
\cite{mag00,takagi2008,takagi2011,elbing2013,muradoglu2014} and thus the bubble distribution in the flow. 
Apparently, the large and deforming 
bubbles, which accumulate close to the inner cylinder \cite{gil13},
 hinder the angular momentum exchange between
boundary layer and bulk by partly blocking the emission of coherent 
structures from the boundary layer towards the bulk and reducing the Reynolds stress,  thus leading
 to drag reduction \cite{lu05,gil13,muradoglu2014, kit05}.  

Our result have  strong bearing on the projected bubble drag reduction in the navel industry.
Not only surfactants, but also ions of the various dissolved salts have 
a strong effect on  coalescence properties of bubbles, either enhancing or suppressing  coalescence
\cite{craig2004}. As seen from our experiments, tests of bubbly
 drag reduction in fresh water facilities will therefore lead to very different results as in the salty
 ocean water. 

Our results however also offer 
opportunities to enhance drag reduction in pipelines transporting liquified  natural gases (LNGs) close
to the boiling point by  adding appropriate surfactants {\it helping}
 coalescence \cite{ike14}.  Going beyond bubbly multiphase flow towards emulsions of e.g.\ oil in water \cite{wong2015}, also here
 the global drag will be strongly affected by the local coalescence behavior of the droplets, thus opening 
 opportunities to 
 influenced the overall drag by the  use of surfactants.

\begin{acknowledgments}
We would like to thank Gert-Wim Bruggert and Martin Bos for their continuous technical support over the years. We acknowledge stimulation discussions with Dennis Bakhuis, and Rodrigo Ezeta Aparicio and Michiel van Limbeek. The work was supported by the Dutch Foundation for Fundamental Research on Matter (FOM) and the Dutch Technology Foundation STW.
\end{acknowledgments}

\bibliographystyle{prsty_allauthors}

\end{document}